\documentstyle[epsf,rotate]{elsart}
\begin{document}
\begin{frontmatter}
\title{Neutrino Propagation in Matter\thanksref{talk}}
\thanks[talk]{Expanded version of a talk at the Slansky Memorial
  Symposium, Los Alamos, May 1998.}

\author{A.B. Balantekin\thanksref{EMAIL}}
\thanks[EMAIL]{Electronic address:
        {\tt baha@nucth.physics.wisc.edu}}

\address{Institute for Nuclear Theory, University of Washington, Box
         351550\\
         Seattle, WA 98195-1550 USA\\
         and\\
         Department of Astronomy, University of Washington, 
          Box 351580\\
       Seattle WA 98195-1580 USA\\
		and\\
        Department of Physics, University of Wisconsin\\
         Madison, Wisconsin 53706 USA\thanksref{add}}
     
\thanks[add]{Permanent Address}

\begin{abstract}
The enhancement of neutrino oscillations in matter is briefly
reviewed. Exact and approximate solutions of the equations 
describing neutrino oscillations in matter are discussed. The role of
stochasticity of the media that the neutrinos propagate through is
elucidated.       
\end{abstract}
\end{frontmatter}

\section{Introduction}

Particle and nuclear physicists devoted an increasingly intensive
effort during the last few decades to searching for evidence of
neutrino mass. Recent announcements by the Superkamiokande
collaboration of the possible oscillation of atmospheric neutrinos
\cite{atm} and very high statistics measurements of the solar
neutrinos \cite{solar} brought us one step closer to understanding the
nature of neutrino mass and mixings. Experiments imply that neutrino
mass is small and the seesaw mechanism \cite{seesaw}, to the
development of which Dick Slansky contributed, is perhaps the simplest
model which leads to a small neutrino mass.

If the neutrinos are massive and different flavors mix they will
oscillate as they propagate in vacuum \cite{ponte}. Dense matter can
significantly amplify neutrino oscillations due to coherent forward
scattering. This behavior is known as the Mikheyev-Smirnov-Wolfenstein
(MSW) effect \cite{msw}.  Matter effects may play an important role in
the solar neutrino problem \cite{recpl,barger,hata}; in transmission
of solar \cite{hata} and atmospheric neutrinos \cite{petpet,newpet}
through the Earth's core; and shock re-heating \cite{shock} and
r-process nucleosynthesis \cite{rproc} in core-collapse supernovae. If
the neutrinos have magnetic moments matter effects may also enhance
spin-flavor precession of neutrinos \cite{spinf}.

The equations of motion for the neutrinos in
the MSW problem can be solved by direct numerical integration, which
must be repeated many times when a broad range of mixing parameters
are considered. This often is not very convenient; consequently
various approximations are widely used. Exact or approximate analytic
results allow a greater understanding of the effects of parameter
changes. The purpose of this article is to present a review of the
solutions of the neutrino propagation equations in matter. Recent
experimental developments and astrophysical implications of the
neutrino mass and mixings are beyond the scope of this article. Very
rapid developments make a medium such as the World Wide Web more
suitable for the former and the latter was recently reviewed elsewhere
\cite{baha1}. Recent experimental developments can be accessed through
the special home page at SPIRES \cite{slac} and theoretical results at
the Institute for Advanced Study \cite{bahcall} and the University of
Pennsylvania \cite{penn}. An assessment of the Superkamiokande solar
neutrino data was recently given by Bahcall, Krastev, and Smirnov
\cite{recpl}. A number of recent reviews cover implications of recent
results for neutrino properties \cite{neupr}.

\section{Outline of the MSW Effect}

The evolution of flavor eigenstates in matter is governed by the
equation \cite{msw,othermsw} 
\begin{equation}
i\hbar \frac{\partial}{\partial x} \left[\begin{array}{cc} \Psi_e(x)
\\ \\ \Psi_{\mu}(x) \end{array}\right] = \left[\begin{array}{cc}
\varphi(x) & \sqrt{\Lambda} \\ \\ \sqrt{\Lambda} & -\varphi(x)
\end{array}\right]
\left[\begin{array}{cc} \Psi_e(x) \\ \\ \Psi_{\mu}(x)
  \end{array}\right]\,, 
\label{1}
\end{equation}
where 
\begin{equation}
  \label{2}
  \varphi(x) = \frac{1}{4 E} \left( 2 \sqrt{2}\ G_F N_e(x) E -  \delta
m^2 \cos{2\theta_v} \right)
\end{equation}
for the mixing of two active neutrino flavors and 
\begin{equation}
  \label{2a}
  \varphi(x) = \frac{1}{4 E} \left( 2 \sqrt{2}\ G_F \left[ N_e(x) -
  \frac{N_n(x)}{2} \right] E -  \delta
m^2 \cos{2\theta_v} \right)
\end{equation}
for the active-sterile mixing. In these equations 
\begin{equation}
  \label{3}
  \sqrt{\Lambda} = \frac{\delta m^2}{4 E}\sin{2\theta_v}, 
\end{equation}
$\delta m^2 \equiv m_2^2 - m_1^2$ is the vacuum
mass-squared splitting, $\theta_v$ is the vacuum mixing angle,  $G_F$
is the Fermi  constant, and $N_e(x)$ and $N_n(x)$ are the number
density of electrons and neutrons respectively  
in the medium.

In a number of cases adiabatic basis greatly simplifies the problem. 
By making the change of basis 
\begin{equation}
  \label{4}
  \left[\begin{array}{cc} \Psi_1(x) \\ \\ \Psi_2(x) \end{array}\right]
= \left[\begin{array}{cc} \cos{\theta(x)} & -\sin{\theta(x)} \\ \\
\sin{\theta(x)} & \cos{\theta(x)}
\end{array}\right]
\left[\begin{array}{cc} \Psi_e(x) \\ \\ \Psi_{\mu}(x)
\end{array}\right]\,,
\end{equation}
the flavor-basis Hamiltonian of Eq.~(\ref{1}) can be instantaneously
diagonalized. The matter mixing angle in Eq.~(\ref{4}) is defined via 
\begin{equation}
\sin{2\theta(x)} = \frac{\sqrt{\Lambda}}{\sqrt{\Lambda + \varphi^2(x)}}
\label{5}
\end{equation}
and
\begin{equation}
\cos{2\theta(x)} =  \frac{-\varphi(x)}{\sqrt{\Lambda +
\varphi^2(x)}}\,.
\label{6}
\end{equation}
In the adiabatic basis the evolution equation takes the form 
\begin{equation}
i\hbar \frac{\partial}{\partial x} \left[\begin{array}{cc} \Psi_1(x)
\\ \\ \Psi_2(x) \end{array}\right] = \left[\begin{array}{cc}
-\sqrt{\Lambda + \varphi^2(x)} & -i \hbar \theta'(x) \\ \\ i \hbar
\theta'(x) & \sqrt{\Lambda + \varphi^2(x)}
\end{array}\right]
\left[\begin{array}{cc} \Psi_1(x) \\ \\ \Psi_2(x)
  \end{array}\right]\,, 
\label{7}
\end{equation}
where prime denotes derivative with respect to $x$.  Since the
$2\times2$ ``Hamiltonian'' in Eq. (\ref{7}) is an element of the
$SU(2)$ algebra, the resulting time-evolution operator is an element
of the $SU(2)$ group. Hence it can be written in the form \cite{dick} 
\begin{equation}
  \label{7aa}
 U =  \left[\begin{array}{cc} \Psi_1(x)&  - \Psi_2^*(x) \\ \\
\Psi_2(x) & \Psi_1^*(x)
\end{array}\right] \,,
\end{equation}
where $\Psi_1(x)$ and $\Psi_2(x)$ are solutions of Eq. (\ref{7}) with
the initial conditions  $\Psi_1(x_i)=1$ and $\Psi_2(x_i)=0$. 
if the matter mixing angle, $\theta(x)$, is changing very slowly
(i.e., adiabatically) its derivatives in Eq.~(\ref{7}) can be set to
zero. In this approximation the ``Hamiltonian'' in the adiabatic basis
is diagonal and the system remains in one of the matter eigenstates. 

To calculate the electron neutrino survival probability  Eq. (\ref{1})
needs to be solved with the initial conditions $\Psi_e=1$ and
$\Psi_{\mu}=0$. Using Eq. (\ref{7aa})  the general solution satisfying
these initial conditions can be written as
\begin{eqnarray}
  \label{7a}
  \Psi_e (x) &=& \cos{\theta(x)} [ \cos{\theta_i} \Psi_1(x) -
\sin{\theta_i} \Psi_2^* (x)] \nonumber \\&+& \sin{\theta(x)} [
  \cos{\theta_i} 
\Psi_2(x) +  \sin{\theta_i} \Psi_1^* (x)],
\end{eqnarray}
where  $\theta_i$ is the initial matter angle. Once the neutrinos
leave the dense matter (e.g. the Sun), the solutions of Eq. (\ref{7})
are particularly simple. Inserting these into Eq. (\ref{7a}) we obtain
the electron neutrino amplitude at a distance $L$ from the solar
surface to be 
\begin{eqnarray}
  \label{7b}
   \Psi_e (L) &=& \cos{\theta_v} [ \cos{\theta_i} \Psi_{1,(S)} -
\sin{\theta_i} \Psi_{2,(S)}^* ] \exp{\left(i\frac{\delta
m^2}{4E}L\right)} \nonumber \\ &+& \sin{\theta_v} [ \cos{\theta_i}
\Psi_{2,(S)} - \sin{\theta_i} \Psi_{1,(S)}^* ]
\exp{\left(-i\frac{\delta m^2}{4E}L\right)},
\end{eqnarray}
where $\Psi_{1,(S)}$ and $\Psi_{2,(S)}$ are the values of  $\Psi_1(x)$
and $\Psi_2(x)$ on the solar surface.  The electron neutrino survival
probability averaged over the detector position, $L$, is then given by 
\begin{eqnarray}
  \label{7c}
  P(\nu_e \rightarrow \nu_e) &=& \langle |\Psi_e (L)|^2 \rangle_L =
\frac{1}{2} + \frac{1}{2} \cos{2\theta_v} \cos{2\theta_i}  \left( 1
- 2 |\Psi_{2,(S)}|^2 \right) \nonumber \\ &-& \frac{1}{2}
\cos{2\theta_v} \sin{2\theta_i} \left( \Psi_{1,(S)} \Psi_{2,(S)} +
\Psi_{1,(S)}^* \Psi_{2,(S)}^* \right)\,.
\end{eqnarray}

If the initial density is rather large, then $\cos{2\theta_i} \sim -1$
and  $\sin{2\theta_i} \sim 0$ and the last term in Eq. (\ref{7c}) is
very  small.  Different neutrinos arriving the detector carry
different phases if they are produced over an extended source. Even if
the initial matter density is not very large, averaging over the
source position makes the last term very small as these phases average
to zero. The completely averaged result for the electron neutrino
survival probability is then given by \cite{parwick} 
\begin{equation}
  \label{7d}
   P(\nu_e \rightarrow \nu_e) = \frac{1}{2} + \frac{1}{2}
\cos{2\theta_v} \langle \cos{2\theta_i} \rangle_{\rm source}  \left( 1
- 2 P_{\rm hop} \right), 
\end{equation}
where the hopping probability is 
\begin{equation}
  \label{7e}
  P_{\rm hop} = |\Psi_{2,(S)}|^2, 
\end{equation}
obtained by solving Eq. (\ref{7}) with the initial conditions
$\Psi_1(x_i)=1$ and $\Psi_2(x_i)=0$. Note that, since in the adiabatic
limit $\Psi_{2,(S)}$ remains to be zero $P_{\rm hop} =0$. 

\section{Exact Solutions}

Exact solutions for the neutrino propagation equations in matter exist
for a limited class of density profiles that satisfy an integrability
condition called shape invariance \cite{coop}. To illustrate this
integrability condition we introduce the operators
\begin{eqnarray}
  \label{9}
  {\hat A}_- &=& i\hbar \frac{\partial}{\partial x} - \varphi(x) \,,
\nonumber \\ {\hat A}_+ &=& i\hbar \frac{\partial}{\partial x} +
\varphi(x) \,.
\end{eqnarray}
Using Eq.~(\ref{9}) Eq. (\ref{1}) takes the form 
\begin{eqnarray}
  \label{10}
  {\hat A}_- \Psi_e(x) &=& \sqrt{\Lambda} \Psi_{\mu}(x)\,, \nonumber
\\ {\hat A}_+ \Psi_{\mu}(x) &=& \sqrt{\Lambda} \Psi_e(x)\,.
\end{eqnarray}
The shape invariance condition  
can be expressed in terms of the operators defined in
Eq. (\ref{9}) \cite{baha2}
\begin{equation}
  \label{13}
  {\hat A}_-(a_1) {\hat A}_+(a_1) = 
  {\hat A}_+(a_2) {\hat A}_-(a_2) + R(a_1). 
\end{equation}
We also introduce a similarity
transformation which formally replaces $a_1$ by $a_2$:
\begin{equation}
  \label{14}
  \hat{T}(a_1) {\cal O}(a_1) \hat{T}^{-1}(a_1) = {\cal O}(a_2).
\end{equation}
The MSW equations take a particularly simple form using the operators
\cite{baha3} 
\begin{eqnarray}
  \label{17}
  \hat{B}_+ &=& \hat{A}_+(a_1) \hat{T} (a_1) \nonumber \\
 \hat{B}_-  &=& \hat{T}^{-1}(a_1) \hat{A}_- (a_1) \,,
\end{eqnarray}
which satisfy the commutation relation: 
\begin{equation}
  \label{18}
  [ \hat{B}_- , \hat{B}_+ ] = R(a_0) ,
\end{equation}
where $a_0$ is defined using the identity 
\begin{equation}
  \label{19}
  R(a_n) = \hat{T}(a_1) R(a_{n-1})\hat{T}^{-1}(a_1),    
\end{equation}
with $n=1$.  
Two additional commutation relations 
\begin{equation}
  \label{20}
   [\hat{B}_+ \hat{B}_- , \hat{B}_+^n ] = (R(a_1)+R(a_2)+ \cdot \cdot 
+ R(a_n))  \hat{B}_+^n , 
\end{equation}
and 
\begin{equation}
  \label{21}
  [\hat{B}_+ \hat{B}_- , \hat{B}_-^{-n} ] = (R(a_1)+R(a_2)+ \cdot \cdot 
+ R(a_n))  \hat{B}_-^{-n} \,,
\end{equation}
can easily be proven by induction. 

Using the operators introduced in Eq. (\ref{17}), Eq. (\ref{1}) can
be rewritten as 
\begin{equation}
  \label{22}
  {\hat B}_+ {\hat B}_- \Psi_e(x) = \Lambda \Psi_e(x).
\end{equation}
Eqs. (\ref{20}) and (\ref{21}) suggest that ${\hat B}_+$ and ${\hat B}_-$
can be used as ladder operators to solve Eq. (\ref{22}). 
Introducing
\begin{equation}
  \label{24}
  \Psi_-^{(0)}  \sim \exp{\left( - i \int \varphi(x;a_1) dx
    \right)}. 
\end{equation}
one observes that 
\begin{equation}
  \label{23}
   {\hat A}_-(a_1) \Psi_-^{(0)} = 0 = {\hat B}_- \Psi_-^{(0)} . 
\end{equation}
If the function 
\begin{equation}
  \label{25}
  f(n) = \sum_{k=1}^{n} R(a_k) 
\end{equation}
can be analytically continued so that the condition
\begin{equation}
  \label{26}
  f(\mu) = \Lambda 
\end{equation}
is satisfied for a particular (in general, complex) value of $\mu$,
then Eq. (\ref{20}) implies that one solution of Eq. (\ref{22}) is 
${\hat B}_+^{\mu} \Psi_-^{(0)}$. Similarly the wavefunction 
\begin{equation}
  \label{29}
   \Psi_+^{(0)}  \sim \exp{\left( + i \int \varphi(x;a_0) dx \right)}\,,
\end{equation}
satisfies
the equation
\begin{equation}
  \label{27}
  {\hat B}_+ \Psi_+^{(0)} = 0. 
\end{equation}
Then a second solution of Eq. (\ref{22}) 
is given by
${\hat B}_-^{-\mu-1} \Psi_+^{(0)}$. Hence for shape invariant electron
densities the exact electron neutrino amplitude can be written as
\cite{baha3} 
\begin{eqnarray}
  \label{30}
  \Psi_e (x) &=& \beta {\hat B}_+^{\mu} \exp{\left( -  i \int
      \varphi(x;a_1) dx \right)} \nonumber \\ &+& 
\gamma {\hat B}_-^{-\mu-1} 
      \exp{\left( + i \int \varphi(x;a_0) dx \right)}, 
\end{eqnarray}
where $\beta$ and $\gamma$ are to be determined using the initial
conditions $\Psi_1(x_i)=1$ and $\Psi_2(x_i)=0$. 

For the linear density profile
\begin{equation}
  \label{41}
  N_e(x) = N_0 - N_0' (x-x_R),
\end{equation}
where $N_0$ is the resonant density: 
\begin{equation}
  \label{42}
  2 \sqrt{2}\ G_F N_0 E =  \delta m^2 \cos{2\theta_v},
\end{equation}
using the technique described above 
we can easily write down the hopping probability 
\begin{equation}
  \label{50c}
  P_{\rm hop} = |\Psi_2 (x_f)|^2 = \exp{(- \pi \Omega)} \,,
\end{equation}
where 
\begin{equation}
  \label{48}
   \Omega = \frac{\delta m^2}{4E}
\frac{\sin^2{2\theta_v}}{\cos{2\theta_v}} \frac{N_0}{N_0'}. 
\end{equation}
This is the standard Landau-Zener result \cite{parwick,wick2}. 

For the exponential density profile
\begin{equation}
  \label{51}
  N_e(x) = N_0 e^{-\alpha(x-x_R)}, 
\end{equation}
where $N_0$ is the resonant density given in Eq. (\ref{42}), 
the hopping probability is \cite{wick3}
\begin{equation}
  \label{62}
  P_{\rm hop} = \frac{e^{-\pi \delta (1 - \cos{2\theta_v})} - e^{-2
\pi \delta}}{1-e^{-2 \pi \delta}}, 
\end{equation}
where we defined
\begin{equation}
  \label{63}
  \delta = \frac{\delta m^2}{2E\alpha}. 
\end{equation}

\section{Supersymmetric Uniform Approximation}

\begin{figure}[t]
\centerline{\hbox{\epsfxsize=3 in \epsfbox[66 56 504 644]{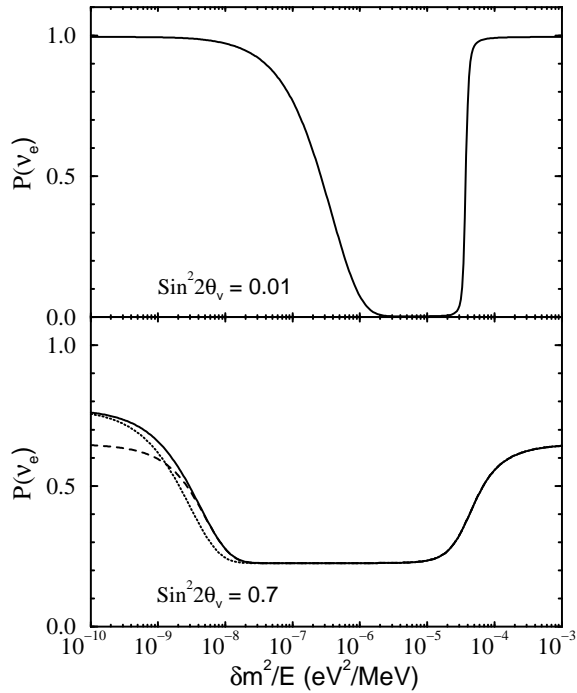}}}
\caption{The electron neutrino survival probability for the Sun
 \cite{baha5}. 
 The solid line is calculated using Eq. ~(\ref{a2}).  The
dashed line is the exact (numerical) result.  The dotted line is the
linear Landau-Zener result.  In the top figure, the lines are
indistinguishable. An exponential density with parameters chosen to
approximate the Sun was used \cite{Bahcall-dens}.}
\label{Fig1}
\vspace{18pt}
\end{figure}

The coupled first-order equations for the flavor-basis wave functions
can be decoupled to yield a second order equation for only the
electron neutrino propagation 
\begin{equation}
-\hbar^2\frac{\partial^2 \Psi_e(x)}{\partial x^2} - \left[\Lambda +
\varphi^2(x) + i\hbar\varphi'(x)\right]\Psi_e(x) = 0.
\label{8}
\end{equation}
The large body of literature on the 
second-order differential equations of mathematical physics
 motivates using a semiclassical approximation for the solutions of
Eq.~(\ref{8}). The standard
semiclassical approximation gives the adiabatic
evolution \cite{baha4}. For a monotonically changing density profile
supersymmetric uniform approximation yields \cite{baha5} 
\begin{eqnarray}
P_{hop} &=& \exp (- \pi \Omega ), \nonumber \\
\Omega &=& \frac{i}{\pi} \frac{\delta m^2}{2 E}
\int^{r_0^*}_{r_0} dr
\left[\zeta^2(r) - 2\zeta(r)\cos{2\theta_v} + 1\right]^{1/2}\,,
\label{a2}
\end{eqnarray}
where ${r_0^*}$ and ${r_0}$ are the turning points (zeros) of the
integrand. In this expression we introduced the scaled density 
\begin{equation}
\zeta(r) = \frac{2\sqrt{2} G_F N_e(r)}{\delta m^2/E}\,,
\label{a3}
\end{equation}
where $N_e$ is the number density of electrons in the medium.  By
analytic continuation, this complex integral is primarily sensitive to
densities near the resonance point. The validity of this approximate
expression is illustrated in Figure \ref{Fig1}.
As this figure illustrates the approximation breaks down in the
extreme non-adiabatic limit (i.e., as $\delta m^2 \rightarrow
0$). Hence it is referred to as the quasi-adiabatic approximation. 

The near-exponential form of the density profile in the Sun 
\cite{Bahcall-dens} motivates an expansion of the electron number
density scale height, $r_s$, in powers of density:
\begin{equation}
 -r_s \equiv \frac{N_e(r)}{N_e'(r)} = \sum_n b_n N_e^n, 
 \label{a4}
\end{equation}
where prime denotes derivative with respect to $r$. In this expression
a minus sign is introduced because we assumed that density profile
decreases a $r$ increases. (For an exponential
density profile, $ N_e \sim e^{-\alpha x}$, only the $n=0$ term is 
present).
\begin{figure}[t]
\centerline{\hbox{\epsfxsize=2.5 in \epsfbox[17 64 539 685]{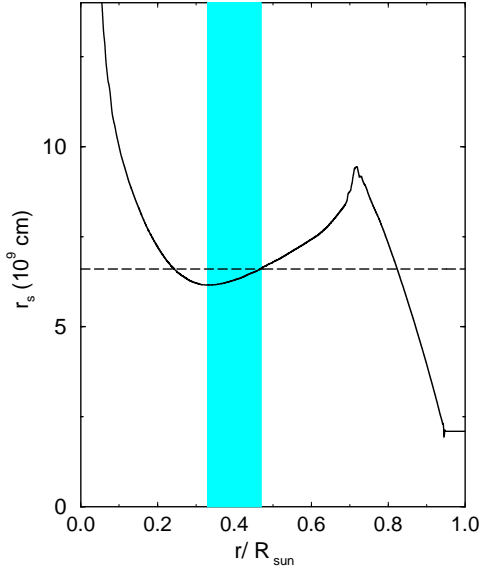}}}
\caption 
{Electron number density scale height (cf. Eq. (\ref{a4})) 
as a function of the radius for
the Sun \cite{Bahcall-dens}. The dashed line is the exponential fit over
the whole Sun. The shaded are indicates where the small angle 
MSW resonance takes
place for neutrinos with energies $5 < E < 15$ MeV.}
\label{Fig1a}
\vspace{18pt}
\end{figure}
To help assess the appropriateness of such an expansion 
the density scale height for the Sun calculated using the Standard
Solar Model density profile is plotted in Figure
\ref{Fig1a}. One observes that there is a significant deviation from a
simple exponential profile over the entire Sun. However the expansion
of Eq.~(\ref{a4}) needs to hold only in the MSW resonance region,
indicated by the shaded area in the figure. Real-time counting
detectors such as Superkamiokande and Sudbury Neutrino Observatory,
which can get information about energy spectra, are sensitive to
neutrinos with energies greater than about 5 MeV. For the small angle
solution ($\sin 2 \theta \sim 0.01$ and $\delta m^2 = 5 \times
10^{-6}$ eV$^2$), the resonance for a 5 MeV
neutrino occurs at about 0.35 
R$_\odot$ and for a 15 MeV neutrino at about 0.45 R$_\odot$ (the shaded
area in the figure). In that region the density profile is 
approximately exponential and one expects that it should be sufficient
to keep only a few terms in the expansion
in Eq.~(\ref{a4}) to represent the density profile of the Standard
Solar Model. 

Inserting the expansion of Eq.~(\ref{a4}) into 
Eq.~(\ref{a2}), and using an integral representation of the Legendre 
functions, one obtains \cite{baha6} 
\begin{eqnarray}
\label{a5}
\Omega &=& -\frac{\delta m^2}{2 E} \left\{ b_0 (1 - \cos{2\theta_v})
\frac{}{} \right. \nonumber \\  &+&   \sum^{\infty}_{n=1}
\left( \frac{ \delta m^2}{2 \sqrt{2} G_F E} \right)^n  \frac{b_n}{2n +
1}  \left. \left[P_{n-1}(\cos{2\theta_v}) -
P_{n+1}(\cos{2\theta_v})\right]\right\}\,, 
\end{eqnarray}
where $P_n$ is the Legendre polynomial of order n.  The $n=0$ term in
Eq.~(\ref{a5}) represents the contribution of the exponential density
profile alone. 
Eq.~(\ref{a5})  directly connects an
expansion of the logarithm of the hopping probability in powers of
$1/E$ to an expansion of the density scale height.  That is, 
 it provides a direct connection between
$N_e(r)$ and $P_\nu(E_\nu)$.  Eq.~(\ref{a5}) provides a
quick and accurate alternative to numerical integration of the MSW
equation for any monotonically-changing 
density profile for a wide range of mixing parameters. 
\begin{figure}[t]
\centerline{\hbox{\epsfxsize=3 in \epsfbox[21 36 566 711]
{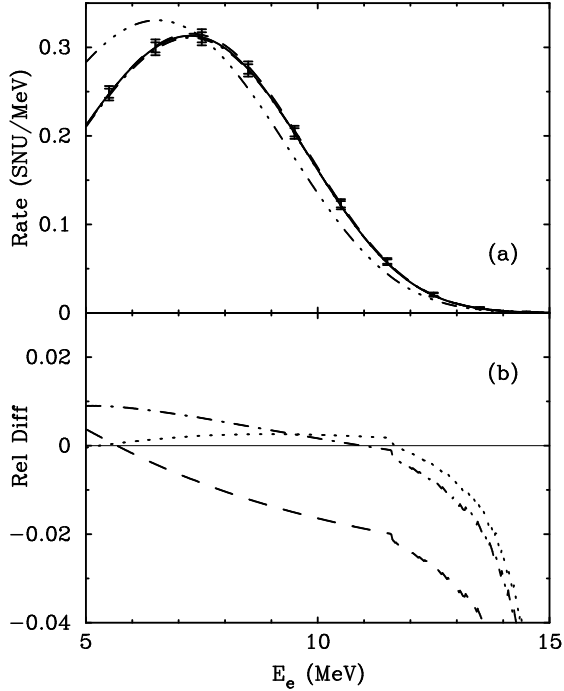}}}
\caption{(a) Spectrum distortion at SNO for the small-angle MSW 
solution ($\delta m^2 \sim 5 \times 10^{-6}$ eV$^2$ and $\sin 2 \theta
\sim 0.01$).  The solid line
is the exact numerical solution.  The dashed, dot-dashed, and dotted
lines result from values of $n$ up to 0, 1, and 2 in Eq.~(\ref{a5}). 
The error bars on the exact numerical result correspond to two and
five years of data collection.  The dot-dot-dot-dashed line is the
spectrum without MSW oscillations, normalized to the same total rate as
with MSW oscillations. Note that on the scale of this figure the $n=1$
and $n=2$ lines are not distinguishable from the exact answer. 
(b) The relative error arising from the use of Eq.~(\ref{a5}).
}
\label{Fig2}
\vspace{18pt}
\end{figure}
The accuracy of the expansion of Eq.~(\ref{a5}) is illustrated in
Figure \ref{Fig2} where the spectrum distortion for the small angle
MSW solution is plotted. In this calculation we used the method 
of Ref.~\cite{snospec} and neglected backgrounds. 
The neutrino-deuterium charged-current cross-sections
were calculated using the code of Bahcall and Lisi \cite{bl}.
One observes that for the Sun, where the
density profile is nearly exponential in the MSW resonance region, the
first two terms in the expansion provide an excellent approximation to
the neutrino survival probability. 

\section{Neutrino Propagation in Stochastic Media}

\indent

In implementing the MSW solution to the solar neutrino problem one
typically assumes that the electron density of the Sun is a
monotonically decreasing function of the distance from the core and
ignores potentially de-cohering effects \cite{sawyer}. To understand
such effects one possibility is to study  
parametric changes in the density or the role of matter currents 
\cite{othernoise}. In this regard, Loreti and Balantekin
\cite{orignoise} considered neutrino propagation in stochastic
media. They studied the situation where the electron density in the
medium has two components, one average component given by the Standard
Solar Model or Supernova Model, etc. and one fluctuating
component. Then the Hamiltonian in Eq. (\ref{1}) takes the form 
\begin{equation}
\hat H =
\left({{-\delta m^2}\over 4E} \cos 2\theta + {1\over \sqrt{2}}
G_F(N_e(r) + N^r_e(r))\right){\sigma_z} + \left({{\delta m^2}\over 4E}
\sin 2\theta \right) {\sigma_x}.  
\end{equation}
where one imposes for
consistency 
\begin{equation}
\langle N^r_e(r)\rangle = 0, 
\end{equation}
 and a two-body
correlation function 
\begin{equation}
\langle N^r_e(r)N^r_e(r^{\prime}) \rangle =
{\beta}^2 \ N_e(r) \ N_e(r^{\prime}) \ \exp(-|r-r^{\prime}|/\tau_c).
\end{equation}
In the calculations of the Wisconsin group the fluctuations are
typically taken to be subject to colored noise, i.e. higher order
correlations 
\begin{equation}
f_{12 \cdots }=\langle N^r_e(r_1)N^r_e(r_2) \cdots
\rangle
\end{equation}
are taken to be 
\begin{equation}
 f_{1234}= f_{12}f_{34} + f_{13}f_{24} +
f_{14}f_{23},
\end{equation}
 and so on.

\begin{figure}[t]
\vspace{8pt} \centerline{\rotate[r]{\epsfxsize=3in
\epsfbox{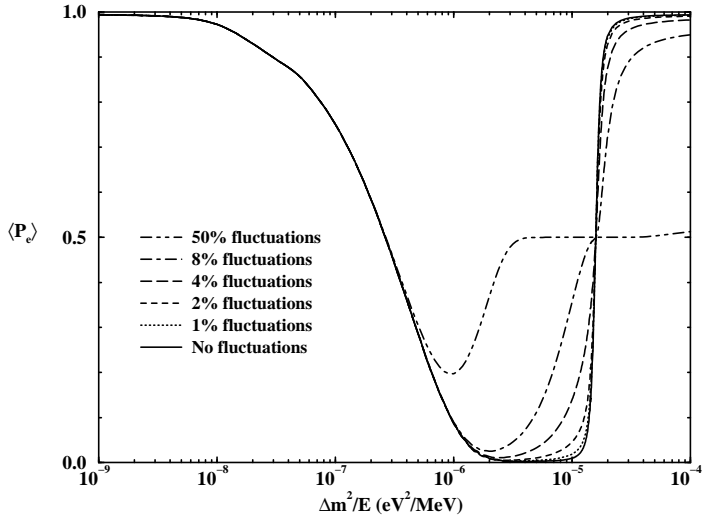}}} 
\caption{
Mean electron neutrino survival probability in the sun with
fluctuations. The average electron density is given by the Standard
Solar Model of Bahcall and Pinsonneault \cite{bp} and $\sin^2 2
\theta=0.01$.} 
\label{Fig3}
\vspace{18pt}
\end{figure}

Mean survival probability for the electron neutrino in the Sun is
shown in Figure \ref{Fig3} \cite{newnoise} where fluctuations are
imposed on the average solar electron density given by the
Bahcall-Pinsonneault model.  

\begin{figure}[t]
\vspace{8pt} \centerline{\rotate[r]{\epsfxsize=3in
\epsfbox{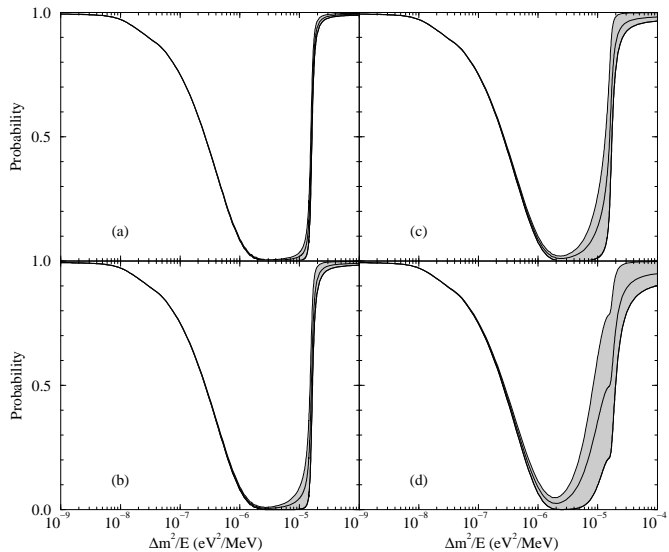}}} 
\caption{ Mean electron neutrino survival probability plus minus
$\sigma$ in the sun with fluctuations. The average electron density is
given by the Standard Solar Model of Bahcall and Pinsonneault and  
$\sin^2 2 \theta=0.01$. Panels (a), (b), (c), and (d) correspond to an
average fluctuation of 1\%, 2\%, 4\%, and 8\% respectively. 
}
\label{Fig4}
\vspace{18pt}
\end{figure}

One notes that for very large
fluctuations complete flavor de-polarization should be achieved,
i.e. the neutrino survival probability is 0.5, the same as the vacuum
oscillation probability for long distances. To illustrate this
behavior the results from the physically unrealistic case of 50\%
fluctuations are shown. Also the effect of the fluctuations is largest
when the neutrino propagation in their absence is adiabatic. This
scenario was applied to the neutrino convection in a core-collapse
supernova where the adiabaticity condition is satisfied
\cite{supernoise}.  Similar results were also obtained by other
authors \cite{nunokawa,burgess,lujan,sendaiguy}. 
It may be possible to test solar
matter density fluctuations at the BOREXINO detector currently under
construction \cite{borex}. Propagation of a
neutrino with a magnetic moment in a random magnetic moment has also
been investigated \cite{orignoise,ranmagnetic}. Also if
the magnetic field in a polarized medium has a domain structure with
different strength and direction in different domains, the
modification of the potential felt by the neutrinos due polarized
electrons will have a random character \cite{polarized}.  Using the
formalism sketched above, it is possible to calculate not only the
mean survival probability, but also the variance, $\sigma$,  of the
fluctuations to get a feeling for the distribution of the survival
probabilities \cite{newnoise} as illustrated in Figure \ref{Fig4}.

In these calculations the correlation length $\tau$ is taken to be
very small, of the order of 10 km., to be consistent with the
helioseismic observations of the sound speed \cite{iki}. In the
opposite limit of very large correlation lengths are very interesting
result is obtained \cite{supernoise}, namely the averaged density
matrix is given as an integral 
\begin{equation}
\lim_{\tau_c\to\infty}\langle\hat \rho(r)\rangle =
{1\over{\sqrt{2\pi \beta^2}}} \int_{-\infty}^{\infty} dx
\exp[{-x^2/(2\beta^2)}]
\hat \rho(r,x),
\end{equation}
reminiscent of the channel-coupling problem in nuclear
physics \cite{takigawa}. Even though this limit is not appropriate to
the solar fluctuations it may be applicable to a number of other
astrophysical situations. 

\begin{ack}
This work was supported in part by the U.S. National Science
Foundation Grant No.\ PHY-9605140 at the University of Wisconsin, and
in part by the University of Wisconsin Research Committee with funds
granted by the Wisconsin Alumni Research Foundation. I thank Institute
for Nuclear Theory and Department of Astronomy  at the University of
Washington for their hospitality and Department of Energy for partial
support during the completion of this work. 
\end{ack}

\end{document}